\documentclass[]{short-report}

\usepackage{lipsum}
\usepackage{amsmath}
\usepackage{mathtools}
\usepackage{bm}
\usepackage{color, colortbl}
\usepackage{authblk}
\usepackage{xr}
\usepackage{graphbox}
\usepackage[export]{adjustbox}
\usepackage[normalem]{ulem}
\usepackage{lipsum}
\usepackage{amsthm}
\usepackage{mdframed}

\theoremstyle{definition}
\newtheorem{exampleinner}{Exemple}[section]

\newmdenv[
  topline=false,
  bottomline=false,
  rightline=false,
  leftline=true,
  linewidth=1pt,
  linecolor=black,
  innerleftmargin=10pt,
  innertopmargin=0pt,
  innerbottommargin=0pt
]{examplebar}

\title{Probability generating functions for epidemiological time series}

\author[1,2]{Mariah C. Boudreau}
\author[2,3]{Guillaume St-Onge}
\affil[1]{The Roux Institute, Northeastern University, Portland, ME 04101, USA}
\affil[2]{Network Science Institute, Northeastern University, Boston, MA 02115, USA}
\affil[3]{School of Electrical Engineering and Computer Science, University of Ottawa, Ottawa, ON K1N 6N5, Canada}

\affil[$\phantom{1}$]{\vspace{-\baselineskip}} 

\date{\today}
\shortauthor{Boudreau et al.}
\shorttitle{}

\let\basetop=\top
\renewcommand{\top}{{}^\basetop \!}

\definecolor{lGray}{gray}{0.95}
\definecolor{lmGray}{gray}{0.90}
\definecolor{mGray}{gray}{0.85}
\definecolor{dGray}{gray}{0.75}

\let\oldalign\align
\let\oldendalign\endalign
\renewenvironment{align}
  {\linenomathNonumbers\oldalign}
  {\oldendalign\endlinenomath}

\begin{document}

\thispagestyle{empty}
\maketitle

\onehalfspacing 
\large 

\section*{Abstract}

Probability generating functions (PGFs) are a fundamental tool in mathematical epidemiology to characterize the stochastic evolution of epidemics via branching process theory.
However, they are rarely used as operational tools in real-time during an outbreak, the main barrier being a lack of a systematic, computational methodology to describe epidemiological time series, such as cases or hospital admissions over time.
Here, we introduce a general and computationally efficient framework for outbreak analytics based on PGFs.
We first develop a method to encode joint distributions of incidences and their noisy measurements by realistic surveillance systems.
Then, by transforming PGFs to cumulant generating functions (CGFs), we enable efficient evaluation of these joint distributions.
We demonstrate the practicality of our approach for Bayesian inference and short-term forecasts, offering a significant computational advantage over simulation-based methods.
Altogether, our framework operationalizes generating functions and paves the way for their broader adoption in computational epidemiology.

\section{Introduction \label{sec:intro}}

Mathematical and computational models are indispensable for predicting outbreak trajectories from real-time observations and informing time-sensitive public health decision-making~\cite{polonsky2019outbreak}.
A variety of modeling frameworks have been developed and deployed as operational tools, such as mechanistic compartmental models~\cite{lipsitch2003transmission,hufnagel2004forecast,arenas2020modeling,chinazzi2024multiscale}, agent-based models~\cite{eubank2004modelling, ferguson2005strategies, aleta2020modelling,kerr2021covasim}, statistical time-series analysis~\cite{helfenstein1986box, box2015time, yang2015accurate}, and semi-mechanistic models based on renewal theory~\cite{feller1941integral, wallinga2004different, cori2013new}. Across these approaches, a persistent tradeoff exists between computational efficiency and the ability to capture heterogeneity in individual-level transmission, such as contact rates shaped by demographic structure and spatial location, and stochastic variation in secondary infections.

Branching process models offer a favorable compromise---they model transmission and disease progression as a stochastic process at the level of individuals, unfolding as a branching tree structure.
This individual-level resolution allows branching processes to capture heterogeneous transmissibility directly through an offspring distribution~\cite{lloyd2005superspreading,meyers2005network}, while remaining efficient to solve using probability generating functions (PGFs).
PGFs streamline the computation of relevant epidemiological quantities, most commonly extinction probabilities, outbreak size distributions, and epidemic thresholds \cite{miller2018primer, southall2023practical, kucharski2015characterizing,  hebert-dufresne2020beyond, allard2023role, curran2025probability}.
They can also characterize the time evolution of contagion in heterogeneously mixed populations, with or without explicit network structure~\cite{noel2009time, allen2022predicting, boudreau2023temporal}, and quantify the probability of disease introduction and detection in metapopulation networks \cite{lahodny2013probability, johansson2014nowcasting, st-onge2025pandemic}.
These results can be used in a broad range of epidemiological tasks, from the evaluation of outbreak risk and control strategies \cite{meyers2003applying, meyers2007contact, toth2015estimates} to parameter inference from the size distributions of transmission chains~\cite{nishiura2012estimating,blumberg2013inference,hebert2021network}.

The aforementioned applications largely target a single summary quantity of the epidemic.
Informing public health decisions in real time requires joint distributions over epidemiological observables across successive reporting periods: noisy, partially ascertained counts, aggregated over fixed time windows.
In this setting, branching process applications often revert to stochastic simulations~\cite{riou2020pattern,worden2019projections}.
Recent work has nonetheless started to bring PGFs closer to epidemic time series in varied contexts.
The PGF of time-varying branching processes has been used to quantify the intrinsic variability of incidence time series and its impact on forecasts~\cite{penn2023intrinsic}.
In phylodynamics, generating functions have enabled the construction of likelihoods combining case counts with genomic data~\cite{zarebski2022computationally,zarebski2025estimating}.
PGFs have also found use in the machine learning literature to perform inference in discrete latent-variable models, including branching processes~\cite{winner2016probabilistic,winner2017exact,zaiser2023exact}.
Additionally, while not relying on PGFs, Gaussian approximations of partially observed branching processes offer a way to enable fast inference with incidence data~\cite{lewis2025gaussian}.

Here, we introduce a general and systematic framework to encode epidemiological time series with generating functions, aiming to broaden the use of PGFs for operational tasks in epidemic modeling.
We first establish our PGF framework through the explicit correspondence between branching process modeling and chain binomial processes---a widely used approach in computational epidemiology to generate stochastic epidemic trajectories in discrete time.
Second, we generalize our encoding approach in two steps: first by tracking incidence at each time step, then the observables derived from it, such as reported cases or hospital admissions.
Third, we provide an efficient method to extract high-dimensional joint distributions of observables, transforming the PGF into a cumulant generating function (CGF) and matching cumulants with a parametric distribution.
We use this approach to demonstrate the capabilities of our framework in key areas of outbreak analytics: short-term forecasting and statistical inference of epidemiological parameters.

\begin{figure}[t]
    \centering
    \includegraphics[width=.9\linewidth]{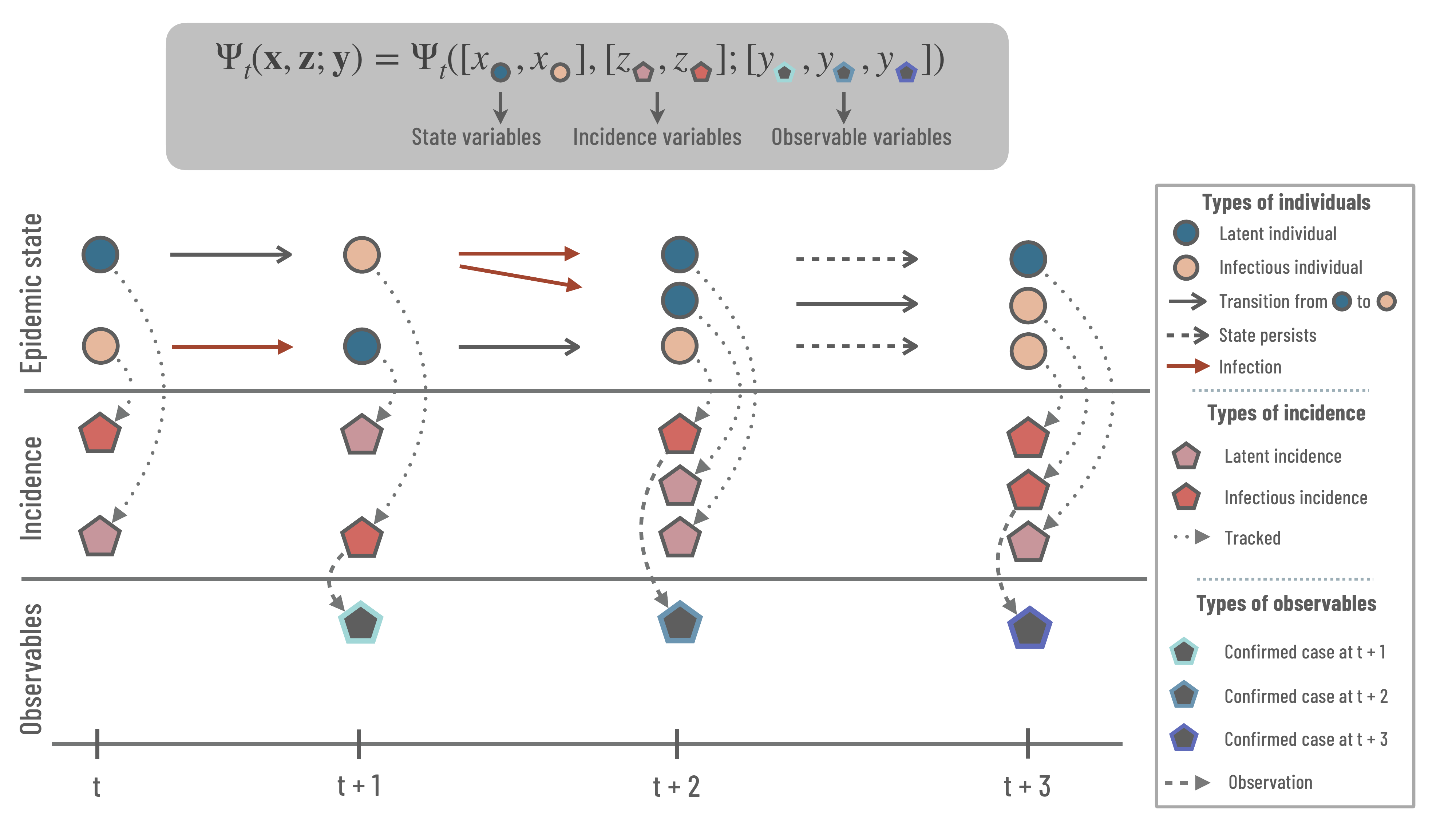}
    \caption{\textbf{Generating function framework for the modeling of epidemiological observables.} We use as an example the Susceptible-Latent-Infectious-Recovered (SLIR) model and cases over time as observables. The top row depicts the underlying epidemic state for two types of individuals, latent and infectious; our SLIR representation does not include susceptible or recovered individuals. The middle row illustrates the incidence of each individual type, directly informed from the epidemic state transitions. The bottom row shows the cases, ascertained with some probability from the infectious incidence. The equation at the top highlights the probability generating function (PGF) at time $t$ targeted by our approach to encode all information about the unfolding process and its measurement. Dummy variables are introduced to track states, incidences, and observables.}
    \label{fig:framework_schematic}
\end{figure}

\section{Probability generating function framework}

We model epidemics as discrete-time multitype branching processes~\cite{harris1963theory}, where each individual of a given type produces a random number of offspring across all types at each time step (see Fig.~\ref{fig:framework_schematic}, top row).
Types, denoted by $\sigma$, are used to partition individuals in the population based on characteristics that influence the evolution of the dynamics, such as disease stages (latent, infectious), demographics (e.g., age), and location for spatial models.
The full state of the epidemic at time $t$ is characterized by $P(\mathbf{n},t)$, the joint probability of having $\mathbf{n} = [n_1, \ldots, n_d]\top$ individuals across $d$ types---we refer to $\mathbf{n}$ thereafter as \emph{state counts}.
Instead of working directly with $P(\mathbf{n},t)$, we encode this distribution through a multivariate PGF,
\begin{align}\label{eq:multitype_pgf}
    \Psi_{t}(\mathbf{x}) \equiv \sum_{\mathbf{n}} P(\mathbf{n},t) \, \mathbf{x}^{\mathbf{n}}, \quad \text{where} \quad \mathbf{x}^{\mathbf{n}} = \prod_{\sigma=1}^{d} x_\sigma^{n_\sigma}.
\end{align}
PGFs are convenient mathematical objects to work with combinatorial structures and to compute distributions efficiently for epidemic dynamics.
Each variable $x_\sigma$ serves as a counting variable for type $\sigma$: monomials $\mathbf{x}^{\mathbf{n}}$ are weighted by the probability $P(\mathbf{n},t)$ of observing state $\mathbf{n}$ at time $t$, making the PGF an alternative representation of the full joint distribution. Indeed, the distribution can always be recovered from $\Psi_t(\mathbf{x})$ via a Fourier integral, or computationally through a discrete Fourier transform~\cite{cavers1978fast}, although this approach does not scale well with the number of types (see supplementary material, text \ref{sec:numerical_inversion_fft}).

To model the evolution of the branching process, we also need to characterize the offspring of each individual of type $\sigma$ at time $t$, which is a random vector $\mathbf{k} = [k_1, \ldots, k_m]\top$ representing counts of individuals across all types. This distribution is encoded as
\begin{align}\label{eq:offspring_pgf_type}
    G_{t,\sigma}(\mathbf{x}) = \sum_{\mathbf{k}} P_\sigma(\mathbf{k},t) \, \mathbf{x}^{\mathbf{k}},
\end{align}
and we collect these into a vector of offspring PGFs, $\mathbf{G}_t(\mathbf{x}) = [G_{t,1}(\mathbf{x}), \ldots, G_{t,d}(\mathbf{x})]\top$.

A fundamental result in branching process theory~\cite{harris1963theory} provides the time evolution of $\Psi_t(\mathbf{x})$ through functional composition,
\begin{align}\label{eq:multitype_pgf_composed}
    \Psi_{t+1}(\mathbf{x}) = \Psi_{t}(\mathbf{G}_{t+1}(\mathbf{x})) = \Psi_{0}(\mathbf{G}_{1}(\mathbf{G}_{2}(...(\mathbf{G}_{t}(\mathbf{G}_{t+1}(\mathbf{x})))...))),
\end{align}
where evaluation proceeds from the innermost function (latest time) outward to the initial condition PGF represented by $\Psi_0(\mathbf{x})$, which encodes $P(\mathbf{n},0)$. For an epidemic seeded with a specific state $\mathbf{n}^*$, the initial PGF is simply
\begin{align}\label{eq:initial_condition}
    \Psi_0(\mathbf{x}) = \mathbf{x}^{\mathbf{n}^*} = \prod_{\sigma=1}^{d} x_\sigma^{n_{\sigma}^*},
\end{align}
but arbitrary distributions $P(\mathbf{n},0)$ can be considered.

In this section, we first show how to construct offspring PGFs for chain binomial processes associated with arbitrary compartmental models. Second, with the aim of modeling epidemic time series, we extend the formalism to track incidence of new types of individuals (middle row of Fig.~\ref{fig:framework_schematic}), then to general epidemiological observables over time (bottom row of Fig.~\ref{fig:framework_schematic}). We complete this section with an analysis characterizing the time at which our PGF model diverges from the chain binomial process.

\subsection{PGF representation of chain binomial processes\label{sec:chain_binom}}

Multitype branching processes can be used to accurately approximate the early stochastic dynamics of discrete-time epidemics in large populations. As a starting point, consider the Susceptible-Infectious-Recovered (SIR) model, formulated as a chain binomial process in a population of size $N$. The stochastic evolution of $S_t$, and $I_t$, the number of susceptible and infectious individuals, respectively, follows
\begin{subequations}
\begin{align}
    S_{t+1} &= S_{t} - \Lambda_t \,  \\
    I_{t+1} &= I_{t} + \Lambda_{t} - \mathrm{Bin}(I_{t},\mu)\,, \label{eq:stochastic_sir_infectious}
\end{align}
\end{subequations}
where
\begin{align}
    \Lambda_{t} = \mathrm{Bin}(S_t, \lambda_t)\,, \quad \text{and} \quad \lambda_t = 1 - (1 - \beta_t / N)^{I_t} \;.
\end{align}
The term $\lambda_t$ represents the force of infection, here the per-time-step probability that a susceptible individual becomes infectious; $\mu$ represents the per-time-step probability an infectious individual recovers; and $\beta_{t}$ is the expected number of contacts an infectious individual makes per time step that would cause infection if the contacted person were susceptible. Contacts are made uniformly at random in the population. One should be cautious in interpreting $\mu$ and $\beta_t$ here as rates---the correspondence to continuous-time dynamics is only valid for time steps that are short relative to the time scales of transmission and recovery.
For large $N$ and sufficiently early in the outbreak, $S_t \approx N$ and $\lambda_t \approx \beta_t I_t /N$, which means
\begin{align}
    \Lambda_t \approx \mathrm{Bin}(N, \beta_t I_t /N) \approx \mathrm{Pois}(\beta_t I_t) \;.
\end{align}
Critically, both variation terms, $\mathrm{Bin}(I_{t},\mu)$ and $\mathrm{Pois}(\beta_t I_t)$, correspond to the sum of $I_t$ \textit{identical and independent} random variables, $\mathrm{Bern(\mu)}$ and $\mathrm{Pois}(\beta_t)$ respectively. From this perspective, infectious individuals produce, independently, a number of offspring at the next time step, which is exactly the definition of a branching process.

We can thus translate the chain binomial SIR model to single-type branching process and introduce the corresponding univariate PGF, $\Psi_{t}(x)$, where $x$ is a counting variable for infectious individuals (only one $\sigma$-type). We simply need to define the offspring PGF for infectious individuals, $G_{t}(x)$. As stated above for the chain binomial, there are two components contributing to infectious individuals offspring, a Bernoulli process for infectious individuals transitioning to the recovered state, $1-\mathrm{Bern}(\mu)$, and a Poisson process producing new infectious individuals, $\mathrm{Pois}(\beta_t)$.
The former process is defined by the PGF,
\begin{equation}
    T(x) = \mu + (1-\mu)x \;,
\end{equation}
which represents the transition from infectious to recovered with probability $\mu$. The latter is given by the PGF,
\begin{equation}
    C_{t}(x) = \exp[\beta_t(x - 1)]\;,
\end{equation}
representing the contagion process. We combine these into a single offspring PGF for infectious individuals as
\begin{align} \label{eq:sir_offspring_pgf}
    G_{t}(x) = T(x)C_{t}(x) =  [\mu + (1-\mu)x] \exp[\beta_t(x-1)] \;,
\end{align}
where we use the product rule for the PGF of a sum of independent random variables~\cite{miller2018primer}.

Let us emphasize here that $G_{t}(x)$ defines the offspring of an infectious individual for the next time step, not the next ``generation" of infectious, which is often the case with branching process approaches describing the transmission tree itself~\cite{miller2018primer}. Our process follows the \textit{physical} time instead---with temporal resolution of time step associated to hours, days, or weeks for instance---, which is crucial to work with surveillance data that have specific time stamps.

\subsubsection*{Generalized disease progression and transmission dynamics}

More realistic descriptions of infectious diseases involve compartmental models with multiple types, representing disease stage, demographics, etc.
Fortunately, the transition--contagion decomposition of the SIR offspring PGF generalizes directly to multitype processes.
In its most general form, an individual of type $\sigma$ transitions to one of many potential types $\sigma'$, each with probability $\mu_{\sigma \to \sigma'}$, a categorical process with PGF
\begin{equation}\label{eq:general_transition_pgf}
    T_{\sigma}(\mathbf{x}) = \sum_{\sigma'}\mu_{\sigma \to \sigma'}x_{\sigma'}.
\end{equation}
Note here, the self-transition probability $\mu_{\sigma \to \sigma}$ may be nonzero, meaning the individual does not change type at the next time step. Similarly, a contagious individual of type $\sigma$ produces on average $\beta_{t,\sigma \to \sigma'}$ new infections of type $\sigma'$ per time step, a Poisson process with PGF
\begin{equation}\label{eq:general_contagion_pgf}
    C_{t,\sigma}(\mathbf{x}) = \exp\left[\sum_{\sigma'} \beta_{t,\sigma \to \sigma'} (x_{\sigma'} - 1) \right].
\end{equation}
Both processes combine in the offspring PGF for type $\sigma$,
\begin{equation}\label{eq:general_offspring_pgf}
    G_{t,\sigma}(\mathbf{x}) = T_{\sigma}(\mathbf{x})C_{t,\sigma}(\mathbf{x}),
\end{equation}
or in vector notation $\mathbf{G}_{t}(\mathbf{x}) = \mathbf{T}(\mathbf{x}) \odot \mathbf{C}_{t}(\mathbf{x})$, by the product rule for PGFs of independent random variables.

We illustrate this construction with an age-structured SLIR (Susceptible-Latent-Infectious-Recovered) model, often the starting point to model epidemiological time series---the incubation period of pathogens creates a delay between the exposure and the beginning of the contagious period~\cite{anderson1991infectious}: latent individuals have been exposed and will become infectious, but cannot yet transmit the disease to others.
Additionally, demographics are a key driver of epidemics and their outcomes: contact patterns~\cite{mistry2021inferring}, as well as hospitalization and fatality rates~\cite{verity2020estimates,prasad2023multimodeling}, are all influenced by the age of individuals, which is why the epidemic modeling community has widely adopted models with age-stratified populations.
We use the index $a \in \{1,\dots, 5\}$ for the five age groups [0-4], [5-19], [20-49], [50-64], and [65+], following established scenario-modeling practice~\cite{prasad2023multimodeling}.

A latent individual becomes infectious with probability $\gamma$ per time step, and an infectious individual recovers with probability $\mu$ per time step; we also define $B_{a,a'}$ as the average number of secondary infection produced in age-group $a'$, per time step, by an infectious individual of age $a$.
The reproduction number for this age-structured SLIR model is $R = \mu^{-1} \rho(\mathbf{B})$, where $\rho(\mathbf{B})$ is the spectral radius of matrix $\mathbf{B}$ with elements $B_{a,a'}$.
We use matrices $\mathbf{B}$ proportional to empirical age-stratified contact matrices in the literature~\cite{mistry2021inferring}.

In our PGF construction, we identify the disease stages by $s \in \{\ell, i\}$, for latent and infectious individuals respectively.
The type of an individual thus corresponds to a pair $\sigma = (s,a)$.
Each type carries a counting variable $x_{s,a}$, and $\mathbf{x}$ collects the ten variables.
While we could add variables to identify them, susceptible and recovered individuals are not tracked as disease stages: both produce no offspring impacting the branching process evolution.

Offspring of latent individuals are produced only from a transition process: a latent individual of age $a$ becomes infectious without changing age, resulting in
\begin{equation}\label{eq:slir_latent_offspring}
    G_{t,\ell,a}(\mathbf{x}) = (1-\gamma)\,x_{\ell,a} + \gamma\, x_{i,a}.
\end{equation}
The offspring PGF for infectious individuals combines transition and contagion, remaining similar in form to Eq.~\eqref{eq:sir_offspring_pgf},
\begin{equation}\label{eq:slir_infectious_offspring}
    G_{t,i,a}(\mathbf{x}) = [\mu + (1-\mu)\,x_{i,a}] \exp\left[\sum_{a'=1}^{5} B_{a,a'}\,(x_{\ell,a'} - 1) \right].
\end{equation}
Since recovered individuals are not tracked, there is no counting variable next to the constant $\mu$.
Note also that variable $x_{\ell,a'}$ is used in the exponential, because infections produce latent individuals.

\begin{figure}[t]
    \centering
    \includegraphics[width=\linewidth]{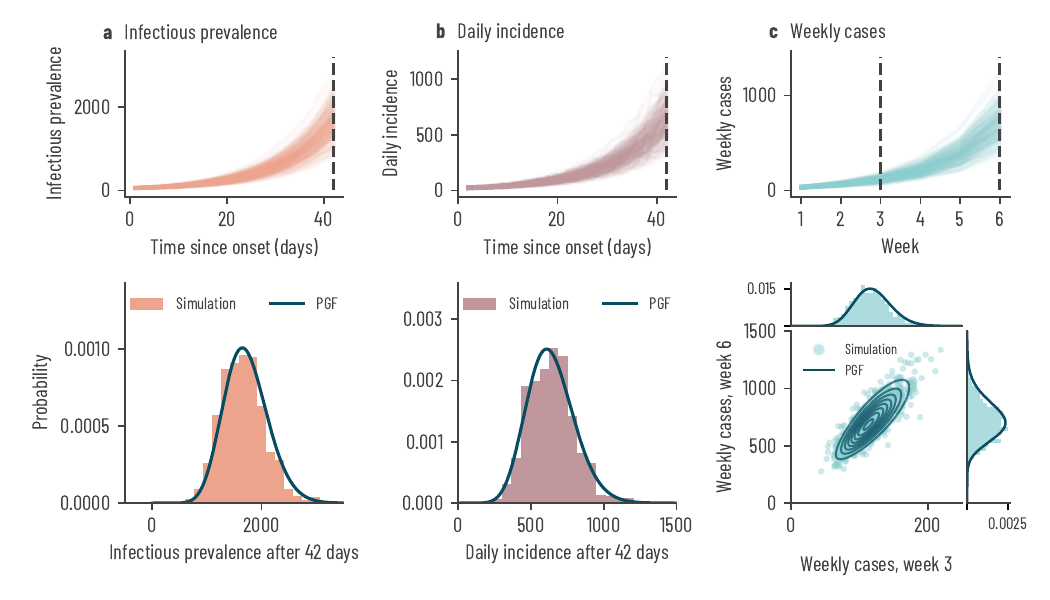}
    \caption{\textbf{Modeling epidemiological observables with probability generating functions.} We compare the distributions computed with our PGF framework with chain-binomial simulations for an SLIR model, using the demographics and age-stratified contact patterns of California, USA \cite{gozzi2025epydemix}. The epidemiological parameters are $R_{0} = 1.5$, a 2-day latent period, and a 3-day infectious period. The initial cluster size is drawn from a Poisson distribution with mean 100, distributed randomly with equal probability to latent and infectious and to each age group according to their proportions in the population. Each column shows a different observed epidemiological quantity with varying observation cadences: \textbf{(a)} infectious prevalence at a single time point, \textbf{(b)} infectious incidence at a single time point, and \textbf{(c)} weekly incident cases with a 20\% case ascertainment rate at two time points. The top row shows 200 \texttt{epydemix} simulations for each observed quantity over 42 days (6 weeks). The bottom row then focuses on the probability distribution from the simulations and PGF for the observed quantity at the corresponding time point(s) of the respective time series (dashed lines).}
    \label{fig:slir_observables}
\end{figure}

In Fig.~\ref{fig:slir_observables}, we illustrate distributions obtained through the PGF $\Psi_t(\mathbf{x})$, which encodes the full joint distribution of the latent and infectious counts,
\begin{equation}\label{eq:slir_full_pgf}
    \Psi_t(\mathbf{x}) = \sum_{\mathbf{n}} P(\mathbf{n},t) \prod_{a=1}^{5} x_{\ell,a}^{\,n_{\ell,a}}\, x_{i,a}^{\,n_{i,a}}\, \quad \text{where } \mathbf{n} = [n_{\ell,1},\dots,n_{\ell,5},n_{i,1},\dots, n_{i,5}]\top \,.
\end{equation}
We consider as initial condition a Poisson distribution for the initial cluster size, with mean $\bar{n}_0$, split with equal probability to latent and infectious and assigned to each age group $a$ with probability $w_a$, corresponding to their proportion in the population
\begin{align}
    \Psi_0(\mathbf{x}) = \exp \left\lbrace \bar{n}_0 \left[ \sum_{a=1}^5 w_a\left(\frac{x_{\ell,a}}{2} + \frac{x_{i,a}}{2} \right) - 1 \right] \right\rbrace \;.
\end{align}

Two substitutions allow extracting quantities of interest from multivariate PGFs. Setting a counting variable to $1$ \emph{marginalizes} over its count, since $x^{n}\big|_{x=1} = 1$; assigning the \emph{same} variable to several types \emph{aggregates} their counts, since a shared base adds exponents, $\prod_a x^{\,n_a} = x^{\sum_a n_a}$. For instance, setting $x_{\ell,a} = 1$ and $x_{i,a} = x_i$ for all $a$ (marginalizing the latent counts and aggregating the infectious counts) yields the PGF for the total number of infectious individuals $n_i \equiv \sum_a n_{i,a}$,
\begin{align}\label{eq:slir_total_infectious_pgf}
    \Psi_t(\mathbf{x})\big|_{x_{\ell,a}=1,\; x_{i,a}=x_i}
    &= \sum_{\mathbf{n}} P(\mathbf{n},t) \prod_{a=1}^{5} 1^{\,n_{\ell,a}}\, x_i^{\,n_{i,a}} \nonumber \\
    &= \sum_{\mathbf{n}} P(\mathbf{n},t)\, x_i^{\,\sum_a n_{i,a}}
    = \sum_{n_i} P(n_i, t)\, x_i^{\,n_i} \,.
\end{align}
Recovering the distribution from this univariate PGF (supplementary material, text~\ref{sec:numerical_inversion_fft}) closely reproduces stochastic simulations, as shown in Fig.~\ref{fig:slir_observables}a.

\subsection{Incidence time series}
So far, $\Psi_t(\mathbf{x})$ only encapsulates the state of the epidemic at time $t$. However, epidemiological quantities measured through surveillance systems are rarely of this nature: most commonly, surveillance data are reported as \textit{incidence} over a defined period of time, e.g., daily cases, weekly hospital admissions, etc. To model these epidemiological observables, we first need a strategy to track the number of \textit{incident} individuals of a particular type $\sigma$, at each time step.

Our approach is to introduce a vector of auxiliary counts $\mathbf{m} = [m_1, \dots, m_d]\top$ where $m_\sigma$ is the number of incident individuals of type $\sigma$ on a given time step---we refer to $\mathbf{m}$ thereafter as \textit{incident counts}. We also define the vector of auxiliary counting variables $\mathbf{z} = [z_1, \dots, z_d]\top$. Our goal is now to evaluate the joint PGF
\begin{align}
    \Psi_{t}(\mathbf{x}, \mathbf{z}) \equiv \sum_{\mathbf{n},\mathbf{m}} P(\mathbf{n},\mathbf{m},t) \mathbf{x}^\mathbf{n} \,\mathbf{z}^{\mathbf{m}}  \quad \text{where} \quad \mathbf{z}^{\mathbf{m}} = \prod_{\sigma=1}^d z_\sigma^{m_\sigma} \;,
\end{align}
encoding both the state of the epidemic and per-time-step incidence.

We can then generalize both the PGF associated with transitions and contagions as follows
\begin{align}
    T_\sigma(\mathbf{x}, \mathbf{z}) &= \mu_{\sigma \to \sigma} x_\sigma + \sum_{\sigma' \neq \sigma} \mu_{\sigma \to \sigma'} x_{\sigma'} z_{\sigma'} \,, \label{eq:ext_transition_other_pgf}\\
    C_{t,\sigma}(\mathbf{x},\mathbf{z}) &= \exp \left[\sum_{\sigma'} \beta_{t,\sigma \to \sigma'} (x_{\sigma'}z_{\sigma'}-1) \right] \,.
\end{align}
Note that we have simply substituted $x_{\sigma'}$ by $x_{\sigma'}z_{\sigma'}$ in terms associated with new incidence of individuals, i.e., resulting from a transition or the generation of secondary infections---crucially, the first term in Eq.~\eqref{eq:ext_transition_other_pgf} do not involve auxiliary variables, since it is for individuals remaining in their current state. The offspring PGF for each type $\sigma$ then simply follows the same structure as before $\mathbf{G}_t(\mathbf{x},\mathbf{z}) = \mathbf{T}(\mathbf{x},\mathbf{z}) \odot \mathbf{C}_t(\mathbf{x},\mathbf{z})$.

However, by introducing the auxiliary variables $\mathbf{z}$, we have implicitly defined new ``types" in our branching process, one additional for each $\sigma$; these auxiliary ``types" thus require an appropriate offspring PGF as well, $\mathbf{F}_t(\mathbf{x},\mathbf{z})$. By definition, we want $\mathbf{z}$ to track incident individuals on a specific time step, which means at the next time step, the count must be 0. The only choice is therefore
\begin{align}
    \mathbf{F}_t(\mathbf{x},\mathbf{z}) = \mathbf{1} \;.
\end{align}
The previous PGF vector acts as a \textit{reset} operation, ensuring incident individuals on a given time step are not carried over to the next. Altogether, we have the generalized recursive evaluation procedure for the joint PGF tracking the state of the epidemic as well as incidence at each time step,
\begin{align}
    \Psi_{t+1}(\mathbf{x}, \mathbf{z}) &= \Psi_t(\mathbf{G}_{t+1}(\mathbf{x},\mathbf{z}), \mathbf{F}_{t+1}(\mathbf{x},\mathbf{z})) = \Psi_t(\mathbf{G}_{t+1}(\mathbf{x},\mathbf{z}), \mathbf{1}) \;. \label{eq:ext_recursion_incidence}
\end{align}

For our age-structured SLIR model, new latent and infectious individuals arise at the contagion and latent-to-infectious transition stages respectively; we track them with incidence variables $z_{\ell,a}$ and $z_{i,a}$, respectively, for each age group. Applying the substitution $x_{\sigma'} \to x_{\sigma'} z_{\sigma'}$ of Eq.~\eqref{eq:ext_transition_other_pgf}, the offspring PGFs become
\begin{subequations}\label{eq:slir_incidence_offspring}
\begin{align}
    G_{t,\ell,a}(\mathbf{x},\mathbf{z}) &= (1-\gamma)\,x_{\ell,a} + \gamma\, x_{i,a}\,z_{i,a}, \\
    G_{t,i,a}(\mathbf{x},\mathbf{z}) &= [\mu + (1-\mu)\,x_{i,a}] \exp\left[\sum_{a'=1}^{5} B_{a,a'}\,(x_{\ell,a'}z_{\ell,a'} - 1) \right].
\end{align}
\end{subequations}
By incorporating these incident individuals and reusing the aggregation--marginalization trick, i.e., $x_{\ell,a} = x_{i,a} = z_{\ell,a} = 1$ and $z_{i,a} = z_i$ for all $a$, we produce probability distributions of daily incident infectious, as shown in Fig.~\ref{fig:slir_observables}b.

\subsection{Generic epidemiological observables}
Equation~\eqref{eq:ext_recursion_incidence} now allows us to encode incident quantities, like new infections, but it is limited to incidence \textit{per time step}; in general, we want a more flexible mechanism to measure incidence over arbitrary periods, and also, incorporate the measurement process itself, like imperfect case ascertainment, hospital admission rates, etc.
In line with our previous strategy, let us introduce another vector of counts $\mathbf{o} = [o_1, \dots, o_r]\top$ for epidemiological observables---thereafter \emph{observable counts}---, as well as their associated vector of counting variables $\mathbf{y} = [y_1, \dots, y_r]\top$. We aim then to characterize the multivariate PGF, encoding state counts, incident counts, and observable counts,
\begin{align}\label{eq:multitype_pgf_observe}
    \Psi_{t}(\mathbf{x} , \mathbf{z}; \mathbf{y}) \equiv \sum_{\mathbf{n}, \mathbf{m}, \mathbf{o}} P(\mathbf{n}, \mathbf{m} , \mathbf{o}, t) \, \mathbf{x}^{\mathbf{n}} \, \mathbf{z}^{\mathbf{m}}\, \mathbf{y}^{\mathbf{o}} \quad \text{where} \quad \mathbf{y}^{\mathbf{o}} = \prod_{\alpha=1}^{r} y_\alpha^{o_\alpha}.
\end{align}

Each specific observable count $o_\alpha$ is a random variable possibly depending on state counts $\mathbf{n}$, and incident counts $\mathbf{m}$ from multiple time steps. For instance, a weekly case incidence at a specific date would depend on new symptomatic infectious individuals in the past seven days. To construct each observable systematically, we introduce two vectors of PGFs: $\mathbf{\Phi}_t(\mathbf{y}) = [\Phi_{t,1}(\mathbf{y}),\dots,\Phi_{t,d}(\mathbf{y})]\top$ where each $\Phi_{t,\sigma}(\mathbf{y})$ encodes the contribution of type $\sigma$ individuals at time $t$ to every observable count $o_\alpha$; similarly, $\mathbf{\Xi}_t(\mathbf{y}) = [\Xi_{t,1}(\mathbf{y}),\dots,\Xi_{t,d}(\mathbf{y})]\top$ where each $\Xi_{t,\sigma}(\mathbf{y})$ encodes the contribution of \textit{incident} type $\sigma$ individuals at time $t$ to every observable count $o_\alpha$. By \textit{contribution}, we mean the probability distribution for the number contributing to an observable count $o_\alpha$. In most instances, it will simply be a binary random variable, i.e., whether or not an incident individual of a certain type contributes to the observable (see example below).
Applying the same logic as before, the generalized recursion including observables is given by
\begin{align}
    \Psi_{t+1}(\mathbf{x}, \mathbf{z}; \mathbf{y})  &= \Psi_t(\mathbf{G}_{t+1}(\mathbf{x} \odot \mathbf{\Phi}_{t+1}(\mathbf{y}),\mathbf{z} \odot \mathbf{\Xi}_{t+1}(\mathbf{y})), \mathbf{1} ; \mathbf{y}) \;. \label{eq:multitype_pgf_composed_ext}
\end{align}

For a well-defined set of observables, there is always a maximum time step $t = T$ which contributes to the observable counts---for instance, if an observable corresponds to weekly cases in a \textit{specific week}, then $T$ would be the last day of the week. Therefore, to characterize observables, we only need to iterate Eq.~\eqref{eq:multitype_pgf_composed_ext} up to $T$.
Moreover, in practice we do not need to extract the unobserved state and incident counts $\mathbf{n}$ and $\mathbf{m}$, we can simplify the PGF using the marginalization trick: replacing all variables $x_{\sigma}$ and $z_{\sigma}$ by 1 automatically marginalizes over the associated counts. Hence, we define the PGF of \textit{observables} as $\Psi_T$, marginalized over the full state and incident counts,
\begin{align}
    \label{eq:observables_pgfs}
\zeta(\mathbf{y}) \equiv \Psi_{T}(\mathbf{1} , \mathbf{1}; \mathbf{y}) = \sum_{\mathbf{n},\mathbf{m}, \mathbf{o}}, P(\mathbf{n}, \mathbf{m}, \mathbf{o}, T) \, \mathbf{1}^{\mathbf{n}} \, \mathbf{1}^{\mathbf{m}}\, \mathbf{y}^{\mathbf{o}} = \sum_{\mathbf{o}} P(\mathbf{o}) \, \mathbf{y}^{\mathbf{o}} \,.
\end{align}
Note that on the right-hand side, we drop the time dependency in $P(\mathbf{o})$ after marginalizing over $\mathbf{n},\mathbf{m}$, since observables are well-defined quantities over a fixed time window, not time-dependent quantities like state or incident counts.
From a hidden Markov model perspective, $\zeta(\mathbf{y})$ completely bypasses the explicit modeling of the hidden states in favor of a representation for the observables only, which streamlines the calculation of probability distributions for concrete epidemiological observables.

We illustrate this procedure with weekly incident cases for our SLIR example.
Each $y_\alpha$ tracks cases for a specific week---collection of 7 time steps---and cases are observed from incident infectious individual with probability $q$.
This is achieved using the following contribution PGF for incident infectious of any age $a$
\begin{equation}\label{eq:slir_observable_contribution}
    \Xi_{t,i,a}(\mathbf{y}) = (1-q) + q\, y_{\lceil t/7 \rceil}\;.
\end{equation}
For instance, we see that for time $t \in {1,\dots,7}$, incident infectious contribution is $(1-q) + q y_1$, accumulating case counts for the first week.
We let $\Phi_{t,i,a}(\mathbf{y}) = 1$ for all $a$, and similarly for contributions from latent state and incident counts, since cases here are derived from incident infectious only.

In Fig.~\ref{fig:slir_observables}c, we see that our approach not only reproduces the marginal distributions for weekly incidence, but also captures the joint distribution over different weeks. Since the vector of observables $\mathbf{o}$ can be of arbitrary length and combine different quantities---cases, hospitalizations, etc.---our framework enables generating function encoding of generic multidimensional epidemiological time series combining different surveillance data streams.

\subsection{Divergence from chain binomial process \label{sec:divergence}}
As described in Sec.~\ref{sec:chain_binom}, we draw an explicit correspondence between chain binomial processes and our PGF framework.
It is therefore important to specify when this correspondence do not hold anymore.
The key distinction between the two approaches lies in the assumptions for population size: the chain binomial considers a finite susceptible population, whereas the PGF framework does not model the depletion of susceptibles.
Saturation effects in the former make the two approaches diverge from one another after some time, which raises the question of when this divergence becomes large enough for the approaches to no longer agree. We introduce two divergence metrics to quantify this point of disagreement, both establishing time horizons for the correspondence of PGFs with chain binomial processes.

\begin{figure}[t]
    \centering
    \includegraphics[width=\linewidth]{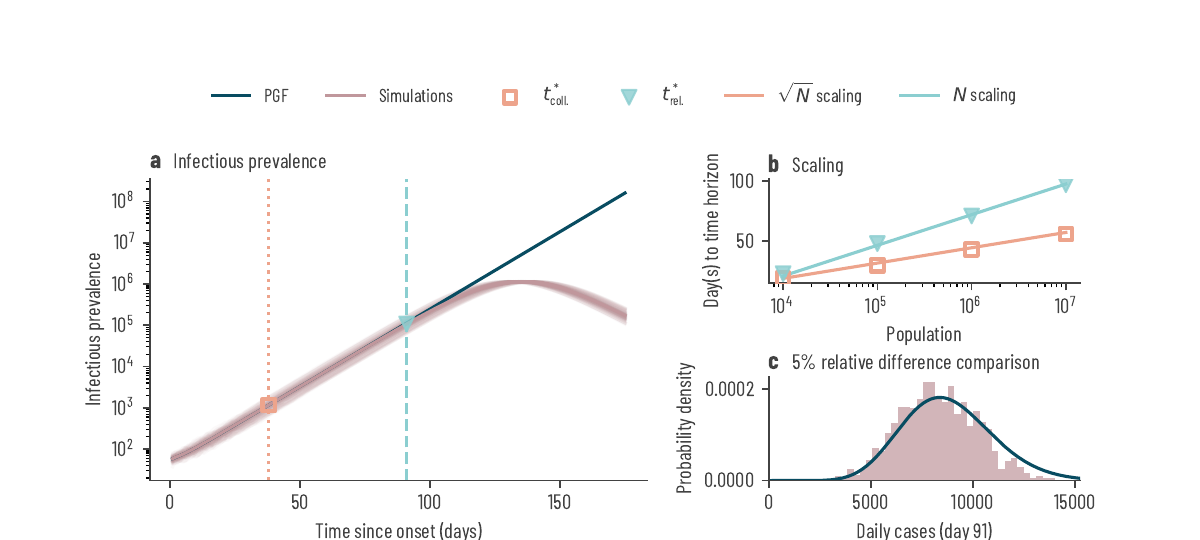}
    \caption{\textbf{Time horizons.} We compare our SLIR model with simulations of the chain binomial process, using the same population, epidemiological parameters, and initial conditions as in Fig.~\ref{fig:slir_observables}. \textbf{(a)} A divergence between simulations and the PGF appears as time exceeds 100 days. We plot two analytical divergence metrics: the collision threshold time horizon, $t^{*}_{\mathrm{coll.}}$ (day 38), and  the 5\% relative difference time horizon, $t^{*}_{\mathrm{rel.}}$ (day 91). \textbf{(b)} For varying synthetic population sizes, both time horizons are plotted as a function of $N$, along with their corresponding scalings. \textbf{(c)} Differences between daily cases with 20\% case ascertainment rate for the simulations and the PGF are shown at the 5\% relative difference time horizon. } 
    \label{fig:divergence}
\end{figure}

Figure \ref{fig:divergence}a illustrates this divergence, showing the point at which each time horizon occurs. The first time horizon, $t^*_{\mathrm{coll.}}$, follows from Ball \emph{et al.}~\cite{ball1995strong}, who show that the branching process approximation for an epidemic model \textit{starts} to breaks down at $t \simeq \frac{1}{2g}\log N$, where $g$ is the growth rate and $N$ is the total population size. This follows from the \emph{collision threshold} of the birthday problem: an infectious individual will attempt to transmit the disease to someone who was already infected in the past when the size of the infected population exceeds $\sqrt{N}$~\cite{ball1995strong}. We use this notion to define $t^*_{\mathrm{coll.}}$, the time at which the infected population surpasses $\sqrt{N}$, given by
\begin{equation}\label{eq:ghost_approx}
    t^*_{\mathrm{coll.}} = \min\left\{ t : \bar{I}_{\mathrm{PGF}}(t) \geq \sqrt{N} \right\},
\end{equation}
where $\bar{I}_{\mathrm{PGF}}$ is the average infectious prevalence extracted from the PGF.

We argue $t^*_{\mathrm{coll.}}$ is a conservative time horizon since it is derived from an asymptomatic argument, rather than an empirical comparison to the results of a chain binomial process. For a sufficiently large $N$, the unfolding predicted by PGFs for $t < t^*_{\mathrm{coll.}}$ is essentially identical to a chain binomial process. A more practical time horizon for prediction---if one can tolerate some level of discrepancy---is the relative difference between $\bar{I}_{\mathrm{PGF}}$ and the average infectious prevalence from simulations, $\bar{I}_{\mathrm{sim.}}$. We focus on the time when a 5\% relative difference occurs, given by
\begin{equation}\label{eq:relative_diff}
    t^*_{\mathrm{rel.}} = \min\left\{ t : \frac{\bar{I}_{\mathrm{PGF}}(t) - \bar{I}_{\mathrm{sim.}}(t)}{\bar{I}_{\mathrm{PGF}}(t)} \geq 0.05 \right\} 
\end{equation}

Figure \ref{fig:divergence}a highlights the large difference between $t^{*}_{\mathrm{coll.}}$ (square marker) at 38 days and $t^{*}_{\mathrm{rel.}}$ (inverted triangle marker) at 91 days. This difference is exacerbated for large populations, since the two time horizon scale differently with $N$, as shown in Fig.~\ref{fig:divergence}b.
Figure \ref{fig:divergence}c is the reason why we call $t^{*}_{\mathrm{rel.}}$ practical: at $t = 91$ days, the PGF still shows very good agreement with the chain binomial provided one accepts an error of the order of 5\%; conversely, $t^{*}_{\mathrm{coll.}}$ would point to a disagreement nearly two months earlier.

Establishing a practical time horizon using $t^{*}_{\mathrm{rel.}}$ is particularly important if the goal is to have predictions match those of a chain binomial process, e.g., if the objective is to use PGF as a surrogate model. One can then choose the level of tolerance based on their needs.
However, in this section we do not want to give the false impression that a chain binomial process is the ``true" epidemic and of the corresponding branching process is the ``wrong" model beyond $t^{*}_{\mathrm{rel.}}$---both integrate strong assumptions about how immunity and population interaction are modeled.
We could compare the predictions of our PGF framework to another model, for example, one that incorporates waning immunity, and find a different time horizon; we could also accomodate in our framework to have time-varying transmission probabilities, hence accounting for the temporal heterogeneity of epidemics and potentially provide a better agreement with data.
Altogether, it is not about stating which model is right or wrong, but of when predictions from different approaches agree.

\section{Application to epidemic prediction and inference}
\label{sec:application_pred_inf}

As illustrated in Fig.~\ref{fig:slir_observables}c, our methodology enables extracting the joint distribution over observables, $P(\mathbf{o})$, which can be used for many applications in computational epidemiology. For instance, if we want to predict future, past, or present observables, $\mathbf{D}'$, based on observed surveillance data, $\mathbf{D}$, then we can use the conditional distribution $P(\mathbf{o}_2 = \mathbf{D}' | \mathbf{o}_1 = \mathbf{D}) \propto P(\mathbf{o}_2 = \mathbf{D}', \mathbf{o}_1 = \mathbf{D})$, where we split observables into two parts, $\mathbf{o}_1$ and $\mathbf{o}_2$.
The joint distribution over observables also directly serves as likelihood $P(\mathbf{o}_1 =\mathbf{D}|\theta)$ for inference purposes, with $\theta$ representing model parameters.

Prediction and inference in epidemiology often require modeling a large number of observables---surveillance observations usually take the form of temporal data streams, hence requiring joint distributions over multiple time points.
Extracting exact, high-dimensional joint probability distribution is impractical, because the number of terms grow exponentially with the number of dimensions.
However, different approaches can be used to approximate full probability distributions or specific values from the PGF $\zeta(\mathbf{y})$~\cite{daniels1954saddlepoint,abate1992numerical,}.
Here, we adopt a simple approach to focus on applications of our generating-function encoding, and leave the exploration of other techniques for future work.
We leverage the connection between PGFs and cumulant generating functions (CGFs)~\cite{gardiner2004handbook,hebert-dufresne2020beyond} to extract first- and second-order cumulants, which we then use for moment matching with parametric distributions.

The CGF of observables is defined as
\begin{align}
    K(\mathbf{s}) = \ln M(\mathbf{s}) = \ln \zeta (\exp(\mathbf{s})) \,,
\end{align}
where $M(\mathbf{s})$ is the moment generating function (MGF). From the CGF, we can recover the mean and covariance structure of observables efficiently through differentiation,
\begin{align}
    \bm{\mu} \equiv E [\mathbf{o} ] = D_\mathbf{s} K(\mathbf{s})\big|_{\mathbf{s} = \mathbf{0}} \quad \mathrm{and} \quad \bm{\Sigma} \equiv \mathrm{Cov}(\mathbf{o}) =  D_\mathbf{s}^2 K(\mathbf{s})\big|_{\mathbf{s} = \mathbf{0}} \,.
\end{align}
This is performed efficiently through automatic differentiation, even for complex epidemic models. Indeed, the composition structure of Eq.~\eqref{eq:multitype_pgf_composed_ext} is analogous to a multilayer neural network, for which automatic differentiation has been optimized and used extensively. In our code base, we make use of \texttt{JAX}~\cite{jax2018github}, which proves to be particularly efficient in performing this task.

We then match the mean and covariance with a parametric distributions to approximate $P(\mathbf{o})$, the distribution of observables. The multivariate normal density function, $P_\mathcal{N}(\mathbf{o}; \bm{\mu}, \bm{\Sigma})$, allows straightforward moment matching, but since epidemiological counts are positive and typically right-skewed, we also considered the multivariate lognormal distribution with a scalar translation parameter $u$, i.e., $\ln(\mathbf{o} - u\,\mathbf{1})$ is multivariate normal (see supplementary material, text \ref{sec:moment_matching}).
We first demonstrate the robustness of our framework with synthetic data, then showcase its application to modeling state-level influenza hospital admissions in the United States.

\subsection{Synthetic data experiments}

\begin{figure}
    \centering
    \includegraphics[width=\linewidth]{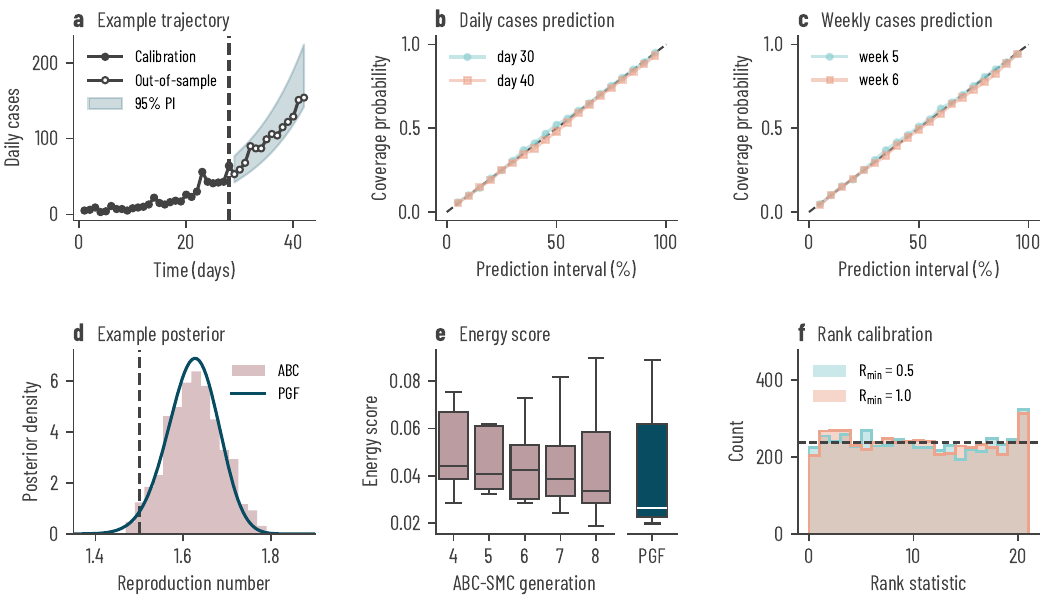}
    \caption{\textbf{Probabilistic forecast and inference on synthetic data.} We consider the same SLIR model, age-structure, and parametrization as Fig.~\ref{fig:slir_observables}. The first 28 days (4 weeks) form the calibration window; days 29--42 (weeks 5--6) form the forecast window. For prediction and inference, we use moment matching with a multivariate lognormal distribution and a translation parameter $u = -10$. \textbf{(a)} Example of stochastic simulation trajectory for daily cases and PGF forecast conditioned on the calibration data. \textbf{(b-c)} Empirical coverage probability compared to the prediction interval indicate well-calibrated forecasts, both for daily and weekly-aggregated case counts. The coverage is evaluated over $n=1000$ synthetic trajectories at two out-of-sample time points each. \textbf{(d-f)} Inference of the reproduction number based on weekly case counts from the calibration window. \textbf{(d)} Example of posterior distribution for PGF and ABC-SMC ($1000$ particles, 8 generations). We use a gamma distribution prior with mean set at the planted value $R^*=1.5$ (dashed line), and a shape parameter of $3$. \textbf{(e)} Energy score obtained over $n = 10$ replicates, using the same gamma prior, for PGF and ABC-SMC using an increasing number of generations. Each box shows the median and interquartile range, the whiskers the 5th and 95th percentiles. The lower the energy score, the better. \textbf{(f)} Simulation-based-calibration of the PGF posterior. The planted parameter $R^*$ is drawn uniformly on $[R_{\min}, 3]$, and a trajectory is simulated. Then we draw $L=20$ samples are from the PGF posterior, evaluated using the same uniform prior. We compile the rank statistic of the of $R^*$ among the sorted posterior samples. The histogram is constructed from $n=5000$ replicates. Under exact posterior sampling, the rank histogram would be uniform on $\{0,\dots,20\}$ (dashed line, shown as a guide to the eye)~\cite{talts2020validating}.}
    \label{fig:synthetic_experiments}
\end{figure}

To test the predictive performance of our PGF framework, we generate synthetic case data from a chain binomial process (Fig.~\ref{fig:synthetic_experiments}a) over a 42-day period, using the same SLIR model as in Fig.~\ref{fig:slir_observables}, and a 20\% case ascertainment. We calibrate our model on the first 28 days of data, $\mathbf{D}$, either using daily or weekly-aggregated cases.
We forecast future cases $\mathbf{D}'$ after the calibration period from the conditional probability distribution $P(\mathbf{D}'|\mathbf{D})$. More specifically, we extract quantiles to produce 20 central prediction intervals (PI), ranging from 5\% to 95\% PI in increment of 5\% (see shaded area in Fig.~\ref{fig:synthetic_experiments}a). We then measure the empirical coverage probability of each prediction interval with the simulated data. We see that the forecasts produced are well-calibrated both at the daily and weekly resolution---the coverage aligns perfectly with the prediction interval, despite the approximation of the conditional distribution $P(\mathbf{D}'|\mathbf{D})$ by a multivariate lognormal. We find similar results with the multivariate normal approximation in Fig.~\ref{fig:synthetic_experiments_normal}, indicating the forecasts are not sensitive to the choice of parametric distribution.

To evaluate inference, we use the same simulation setting to generate synthetic data, focusing on weekly cases, and this time we vary the reproduction number with planted values $R^* \in [0.5, 3]$.
We perform Bayesian inference of the reproduction number $R$ by evaluating the approximate posterior
\begin{align}
    P(R|\mathbf{D}) \propto P(\mathbf{D}|R) P(R) \;, \label{eq:posterior_repr_number}
\end{align}
where $P(\mathbf{D}|R)$ is the approximate likelihood evaluated from our chosen parametric distribution, and $P(R)$ is a prior density distribution.
Since we only infer one parameter for this experiment ($\theta = R$), we simply discretize and evaluate the unnormalized posterior in Eq.~\eqref{eq:posterior_repr_number} on a grid with step-size $\mathrm{d}R$, then normalize the approximate posterior density.

We compare our inference against an approximate Bayesian computation approach coupled with a sequential Monte Carlo sampler (ABC-SMC), a standard simulation-based inference technique~\cite{gozzi2025epydemix, minter2019approximate}.
In all our ABC-SMC inference experiments, we use the root mean square error as the distance between observed and simulated data, an adaptive tolerance schedule in which the tolerance at each generation is the median distance of the particles accepted in the previous generation, and a Gaussian random walk kernel.
One such example of posterior density and samples are illustrated in Fig.~\ref{fig:synthetic_experiments}d, with good agreement between the approaches.
Let us emphasize the computational efficiency of PGF inference relative to ABC-SMC in attaining similar posterior distributions (1000 particles and 8 generations for the SMC sampler).
On an Apple M1 Max, ABC-SMC took about 3.5 minutes (serial execution) and PGF inference about 0.3 seconds, including \texttt{JAX}'s just-in-time compilation.
In Fig.~\ref{fig:synthetic_experiments}e, we further quantify the performance of both approaches for parameter recovery by measuring the energy score~\cite{gneiting2007strictly,bay2025evaluation}. A lower score indicates better performance. Although the score value fluctuates across simulated data realizations, we find a lower median value with the PGF framework compared to ABC-SMC; we could potentially reach lower values by increasing the number of generations for SMC, but at the expense of longer computation time.

Additionally, to diagnose the approximate posterior distribution recovered from our PGF framework, we perform a simulation-based-calibration inspired from Ref.~\cite{talts2020validating}.
We sample $R^*$ from a prior density $P(R)$, simulate case data $\mathbf{D}$, then sample from our posterior $P(R|\mathbf{D})$. Under exact posterior sampling, the rank statistics of $R^*$ should be uniformly distributed among the posterior samples, and deviations from a uniform distribution are indicative of biases or issues with the posterior shape.
We recover rank statistics that are relatively uniform in Fig.~\ref{fig:synthetic_experiments}f, with slightly higher densities toward the ends, especially near the upper bound. We attribute this deviation mostly to the shape of the approximate posterior not capturing exactly the skewness of the true posterior density (see supplementary material, text~\ref{sec:add_synthetic_experiments}).

In supplementary material, text~\ref{sec:add_synthetic_experiments}, we explore more broadly the impact on inference of varying the parametric distributions (multivariate normal versus lognormal, and their hyperparameters) approximating the likelihood.
Contrarily to forecasts, we observe more sensitivity on the performance and calibration, especially when considering daily instead of weekly cases data, in part because the approximating density distributions we use do not represent low case counts well.
Other means of extracting the joint observables distribution should therefore be privileged in such a context, or consider sufficient temporal aggregation to mitigate low count effects.

\subsection{Modeling influenza hospital admission}
\label{sec:influenza_hospital_modeling}

\begin{figure}[t]
    \centering
    \includegraphics[width=\linewidth]{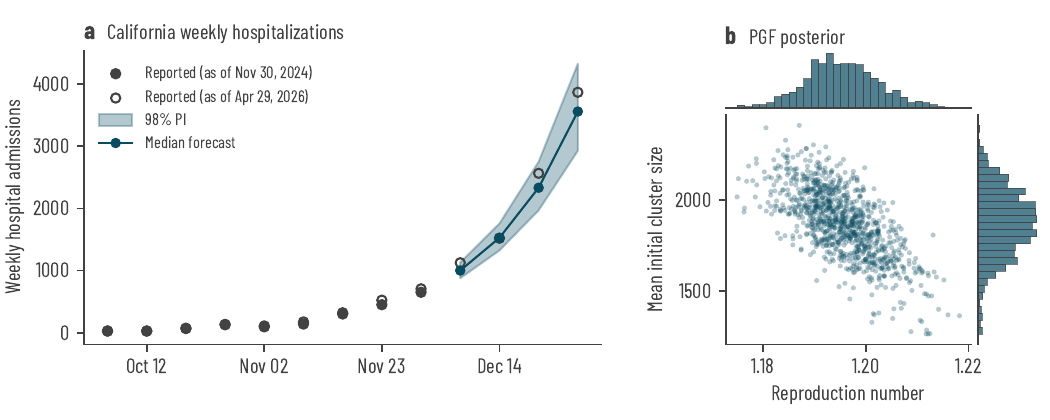}
    \caption{\textbf{Inference and forecast for influenza hospital admissions.} We use an extended SLIR model with pre-hospitalized and hospitalized compartments. The latent and infectious periods are 1.5 days, the probability of an infectious being asymptomatic is 45\%, which leads to a 50\% reduction in transmission. We use California demographics and age-stratified contact patterns~\cite{gozzi2025epydemix}. The probability a symptomatic individual is hospitalized and the pre-hospitalized period are both age-dependent and compiled in Table~\ref{tab:flu_age_params} and informed from Ref.~\cite{prasad2023multimodeling}. For prediction and inference, we use moment matching with a multivariate lognormal distribution and a translation parameter $u = -10$. \textbf{(a)} The black markers represent the calibration data, i.e., weekly hospital admissions from Oct. 5 to Nov. 30, 2024, as reported at the latter date; the white markers are out-of-sample target data for prediction, from Dec. 5 to Dec. 28, 2024, as reported as of April 29, 2026. \textbf{(b)} Parameter inference for the reproduction number and the mean initial cluster size based on the calibration data. We used Hamiltonian Monte Carlo to collect 1000 samples.
    }
    \label{fig:influenza_hosp_modeling}
\end{figure}

The real-time prediction of future infectious disease outcomes, such as cases, hospital admissions, and deaths is an active research area in computational epidemiology, with many ongoing multimodeling efforts~\cite{cramer2022evaluation,howerton2023evaluation,mathis2024evaluation,gozzi2026performance}.
As a proof of concept, we implemented a PGF model of influenza hospital admission to perform inference and forecast.
We build on the SLIR model in previous sections, adding two disease stages following an infection: pre-hospitalized and hospitalized. Only a certain fraction of infectious transition to the pre-hospitalized and hospitalized stages; the pre-hospitalized stage is introduced to set more realistic delays between symptom onset and hospital admissions. Additionally, infectious individuals can be asymptomatic, in which case their transmission is reduced. 
See supplementary material, text~\ref{sec:influenza_model} for detailed model specification.

In Fig.~\ref{fig:influenza_hosp_modeling}, we present an application of our approach to influenza hospitalizations in California in the fall and early winter of 2024.
Since our current model do not account for the depletion of susceptible individuals in the population, we restrict our analysis to the beginning of the influenza season.
With our framework, we define observables for weekly incident hospitalizations, the aggregation level reported to CDC’s National Health Safety Network (NHSN)~\cite{cdcnhsn2026}.
Two parameters of the model are inferred: the effective reproduction number $R$ and the mean initial cluster size at the beginning of the season (here $t = 0$ is fixed to Sept. 29, 2024).
We assume the initial cluster size is Poisson distributed, with equal distribution between latent and infectious, and assignment proportional to each age-group relative frequency.
Parameters $\theta^{(j)}$ are sampled through Hamiltonian Monte-Carlo~\cite{gelman2013bayesian} using data on or prior to Nov. 30, 2024, then we predict weekly hospital admissions for the period of Dec. 7 to Dec. 28, 2024. The forecasts correspond to the aggregation of conditional probability distributions on the inferred parameters
\begin{align}
    P(\mathbf{D}'|\mathbf{D}) \approx \frac{1}{M}\sum_{j=1}^M P(\mathbf{D}'|\mathbf{D}, \theta^{(j)}) \;.
\end{align}
Results of the inference are presented in Fig.~\ref{fig:influenza_hosp_modeling}b, where we obtain a median reproduction number of 1.20, consistent with typical values for seasonal influenza, and a median mean initial cluster size of 1877.
Moreover, the forecasts of our model in Fig.~\ref{fig:influenza_hosp_modeling}a capture the rapid growth of hospital admissions observed at the end of December that year well. As of April 2026, reported hospitalizations for the 4 predicted weeks all fall within the 98\% PI predicted by our approach.

In Fig.~\ref{fig:influenza_hosp_modeling_supp}, we present additional inferences and forecasts for other states during the same time period.
Our forecasts capture the trend in some cases, but fail to identify the onset of the epidemic in others---additionally, the prediction intervals produced are generally too narrow, a sign of model overconfidence~\cite{bracher2021evaluating}.
While our model integrates realistic mechanistic features, like age-stratified transmission patterns and age-dependent disease progression informed from literature, it remains a simplification of the real transmission dynamics, ignoring behavior, seasonality, and detailed population structure to name a few.
Forecasting methods are broadly hindered by model misspecification, limited surveillance data, and fundamental entropy barrier~\cite{scarpino2019predictability}.
This is why a single model often fails to accurately and consistently predict the evolution of epidemics across different locations~\cite{cramer2022evaluation} supporting the use of ensemble methods leveraging the wisdom of the crowd.
The value of this approach is therefore not in replacing current models, but to streamline calculations for complex mechanistic models relying on chain binomial processes, as highlighted in previous section.

\section{Discussion\label{sec:discussion}}

We introduced a mathematical framework to the modeling of epidemiological time series via branching process theory and PGFs.
It allows us to systematically extract joint probability distributions for general outcome of oubreaks over time, derived from the state of the epidemic or its incidence in the population, like cases or hospital admissions.
Being fully analytical, our framework enable efficient computation for inference and prediction of epidemics evolution, two key applications of outbreak analytics.
We demonstrated the operational practicality of our framework through both synthetic experiments and an application to the modeling of influenza hospital admission.

Here, we limited ourselves to age-stratified epidemic models for simplicity, but the approach is straightforward to generalize to integrate more sociodemographic and socioeconomic features. It can also describe spatial dynamics through metapopulation approaches~\cite{hanski1999metapopulation,pastor-satorras2015epidemic}, enabling the modeling of importation and detection of pathogens from travelers~\cite{st-onge2025pandemic}. Future work should formalize the encoding and extraction of such time series within our PGF framework.

When faced with sufficiently high-dimensional observables, we leveraged the CGF to approximate the underlying the joint probability distribution over observables.
Since we limited ourselves to second-order moment matching, the resulting predictions and inferences performance should be similar, in principle, to techniques like Kalman Filtering and linear noise approximations~\cite{shaman2012forecasting,fintzi2022lna,gardiner2004handbook,lewis2025gaussian}.
However, let us highlight that the multivariate PGF encapsulates the full discrete stochastic process.
Hence, any improvement to extracting the underlying distribution---while remaining computationally feasible---would enhance the predictions of this framework.

While we limited our inference experiments to a limited number of parameters, our framework is general and readily applicable to inference tasks in higher dimension. For instance, time-varying reproduction number estimation~\cite{gostic2020practical,nash2022real,milando2026vision,} is an active area in the epidemic modeling community.
In this context, semi-mechanistic renewal models are most often used, which share many of the assumptions of branching process and PGF evaluation, but the set of tools and methods employed are different.
A thorough assessment of PGF inference, benchmarked against current methods and software, would help clarify where PGF modeling stands relative to other approaches in computational epidemiology.

Finally, we foresee many applications and future avenues for the improvement of PGF-based forecasts.
Predictions in the early growth phase of an epidemic or at the beginning of an influenza season could leverage PGF models similar to the one used here.
Posterior inference could be used to seed or constrain the parameter space exploration performed by complex simulation-based models.
Another direction is to to extend the applicability of our approach past the initial growth phase by developing explicit modeling of immunity in the population.

\section*{Acknowledgments}

We thank Alessandro Vespignani for helpful comments on the manuscript.
This work was in part supported by the Gates Foundation (no. INV-058220 and INV-094330). Under the grant conditions of the Foundation, a Creative Commons Attribution 4.0 License has already been assigned to the Author Accepted Manuscript version that might arise from this submission.
We acknowledge support from cooperative agreement no. CDC-RFA-FT-23-0069 from the CDC’s Center for Forecasting and Outbreak Analytics.
GS acknowledges support from the U.S. National Science Foundation award no. 2619828.
The findings and conclusions in this study are those of the authors and do not necessarily represent the official position of the funding agencies, the CDC or the US Department of Health and Human Services.

\printbibliography

\clearpage
\appendix

\setcounter{figure}{0}
\setcounter{table}{0}
\setcounter{section}{0}
\setcounter{equation}{0}

\renewcommand{\thefigure}{S\arabic{figure}}
\renewcommand{\thetable}{S\arabic{table}}
\renewcommand{\thesection}{S\arabic{section}}
\renewcommand{\theequation}{S\arabic{equation}}

\begin{flushleft}\sffamily
  {\LARGE\textbf{Supplementary material}}\par\medskip
  {\large Probability generating functions for epidemiological time series}\par\medskip
  \textsl{Mariah C. Boudreau and Guillaume St-Onge}
\end{flushleft}
\vspace{1em}

\section{Moment matching}
\label{sec:moment_matching}

From the CGF, we can extract the mean vector $\bm{\mu}_\theta \equiv \bm{\mu}$ and the covariance matrix $\bm{\Sigma}_\theta \equiv \bm{\Sigma}$ over the observables $\mathbf{o} = [o_1, \dots, o_r]\top$. Moment matching is straightforward if we assume $\mathbf{o} \sim \mathcal{N}(\bm{\mu},\bm{\Sigma})$, i.e., we approximate the distribution of observables as a multivariate normal (MVN). For statistical inference, we use as likelihood the probability density function of the MVN, $\mathcal{L}(\theta;\mathbf{D}) = P_\mathcal{N}(\mathbf{D}| \bm{\mu},\bm{\Sigma})$, where $\mathbf{D}$ is the observed data.
For forecasting $\mathbf{D}'$ based on observed data $\mathbf{D}$, we directly use the conditional distribution $P_\mathcal{N}(\mathbf{D}'|\mathbf{D},\bm{\mu},\bm{\Sigma})$, which is obtained through standard Gaussian conditioning formulas.

However, epidemiological data are most often nonnegative, discrete, and with right-skewed marginals, at odds with the MVN. Moment matching with discrete distributions is challenging, but a natural alternative to the MVN is the \textit{translated multivariate lognormal distribution}, i.e., we assume
\begin{align}
    \mathbf{o} = e^{\mathbf{v}} + u\,\mathbf{1}, \qquad
    \mathbf{v} \sim \mathcal{N} \left(\hat{\bm{\mu}},\hat{\bm{\Sigma}}\right)\;.
    \label{eq:surrogate_lognormal}
\end{align}
The exponential is applied component wise, $\mathbf{1}$ is a vector of $r$ ones, and $u \in \mathbb{R}$ is a fixed translation parameter. The resulting observable support is $(u,\infty)^r$. We usually take $u < 0$ to accommodate zero counts in the data: the support is then not strictly nonnegative, but it remains bounded from below. Additionally, the resulting marginal distributions are naturally right-skewed, more in line with epidemiological data.

To perform moment matching, we determine the latent normal
parameters $(\hat{\bm{\mu}}, \hat{\bm{\Sigma}})$ from
$(\bm{\mu}, \bm{\Sigma}, u)$ by imposing that the first two cumulants match
\begin{align}
    \mathrm{E}[\mathbf{o}] = \bm{\mu},
    \qquad
    \mathrm{Cov}(\mathbf{o}) = \bm{\Sigma}.
    \label{eq:mm_condition}
\end{align}

To start, the first two cumulants of the translated multivariate lognormal are
\begin{subequations}
\begin{align}
    \mu_\alpha &= e^{\hat\mu_\alpha + \hat\Sigma_{\alpha\alpha}/2} + u,
    \label{eq:target_mean} \\
    \Sigma_{\alpha\alpha'} &= e^{\hat\mu_\alpha + \hat\mu_{\alpha'} + (\hat\Sigma_{\alpha\alpha} + \hat\Sigma_{\alpha'\alpha'})/2}
        \left(e^{\hat\Sigma_{\alpha\alpha'}} - 1\right).
    \label{eq:target_cov}
\end{align}
\end{subequations}
Introducing the shifted target mean $\tilde{\bm{\mu}} = \bm{\mu} - u\,\mathbf{1}$, the covariance condition simplifies to
\begin{align}
    \Sigma_{\alpha\alpha'} = \tilde\mu_\alpha\,\tilde\mu_{\alpha'}\left(e^{\hat\Sigma_{\alpha\alpha'}} - 1\right).
    \label{eq:target_cov_2}
\end{align}
Inverting yields the latent covariance,
\begin{align}
    \hat\Sigma_{\alpha\alpha'} = \log \left(1 + \frac{\Sigma_{\alpha\alpha'}}{\tilde\mu_\alpha\,\tilde\mu_{\alpha'}}\right).
    \label{eq:latent_cov}
\end{align}
We then use this result to solve for the latent mean by inverting
\begin{align}
    \tilde\mu_\alpha = e^{\hat{\mu}_\alpha + \hat\Sigma_{\alpha\alpha}/2},
    \label{eq:shifted_mean}
\end{align}
which gives
\begin{align}
    \hat\mu_\alpha = \log\tilde{\mu}_\alpha - \frac{1}{2}\hat{\Sigma}_{\alpha\alpha}
    = \log\tilde{\mu}_\alpha - \frac{1}{2}\log \left(1 + \frac{\Sigma_{\alpha\alpha}}{\tilde\mu_\alpha^{2}}\right).
    \label{eq:latent_mean}
\end{align}

With the transformations for latent normal parameters in hand, forecast and inference are straightforward, since we can reuse the MVN properties after transforming observed data $\mathbf{D}$ into the latent space
\begin{align}
    \hat{\mathbf{D}} = \log \left( \mathbf{D} - u\,\mathbf{1} \right) \;.
\end{align}
Indeed, $\hat{\mathbf{D}}$ lives in the same space as $\mathbf{v}$, which is an MVN. We can thus reuse the standard Gaussian conditioning formulas to obtain the conditional multivariate lognormal distribution, from which we make predictions for $\mathbf{D}'$. For statistical inference, the likelihood is
\begin{align}
    \mathcal{L}(\theta;\mathbf{D}) = P_\mathcal{N}(\hat{\mathbf{D}}| \hat{\bm{\mu}},\hat{\bm{\Sigma}}) \, \prod_{\alpha=1}^r e^{-\hat{D}_\alpha} \;,
\end{align}
with the product on the right accounting for the Jacobian of the transformation $\mathbf{D} \mapsto \hat{\mathbf{D}}$.

Let us note that the matrix $\hat{\bm{\Sigma}}$ produced by Eq.  \eqref{eq:latent_cov} is symmetric by construction, but it is not guaranteed to be positive definite for arbitrary target cumulants $(\bm{\mu}, \bm{\Sigma})$.
The translated-lognormal family does not span every admissible pair $(\bm{\mu}, \bm{\Sigma})$. In practice, however, we have found the multivariate lognormal distribution to be widely applicable to modeling epidemiological time series data with PGFs.

\section{Additional synthetic data experiments}
\label{sec:add_synthetic_experiments}

\begin{figure}[t]
    \centering
    \includegraphics[width=\linewidth]{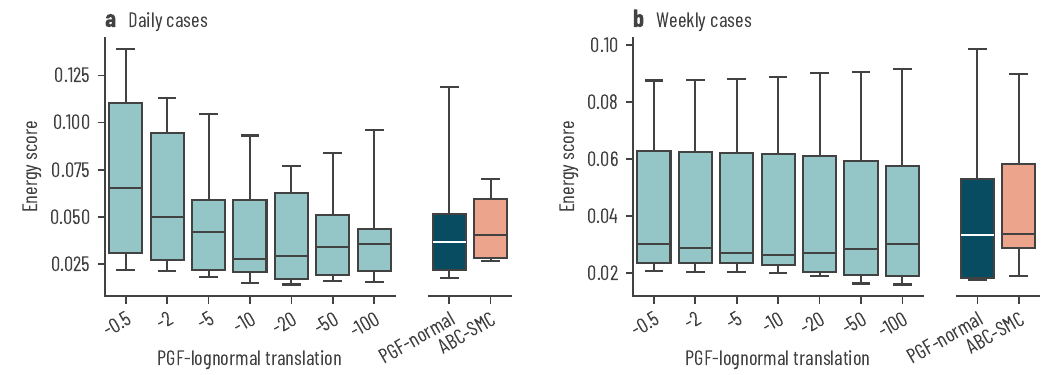}
    \caption{\textbf{Sensitivity analysis of the lognormal moment matching for inference.} Synthetic data and inference settings are identical as in Fig.~\ref{fig:synthetic_experiments}e. The PGF likelihood on $R$ is computed via a lognormal distribution with translation  $u\in\{-0.5,-2,-5,-10,-20,-50,-100\}$, and its energy score is reported. For comparison, we show results for PGF with a normal likelihood and ABC-SMC with 1000 particles and 8 generations. \textbf{(a)} Daily-case calibration. \textbf{(b)} Weekly-case calibration.}
    \label{fig:synthetic_experiments_sensitivity}
\end{figure}

\begin{figure}[t]
    \centering
    \includegraphics[width=\linewidth]{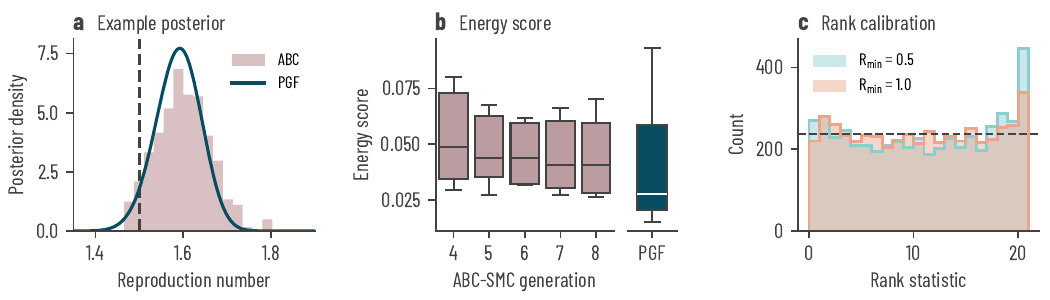}
    \caption{\textbf{Posterior inference on the reproduction number from daily-case calibration}.  Synthetic data and inference settings are identical as in Fig.~\ref{fig:synthetic_experiments}(d-f), but we use daily case counts as data.}
    \label{fig:synthetic_experiments_daily}
\end{figure}

\begin{figure}[t]
    \centering
    \includegraphics[width=\linewidth]{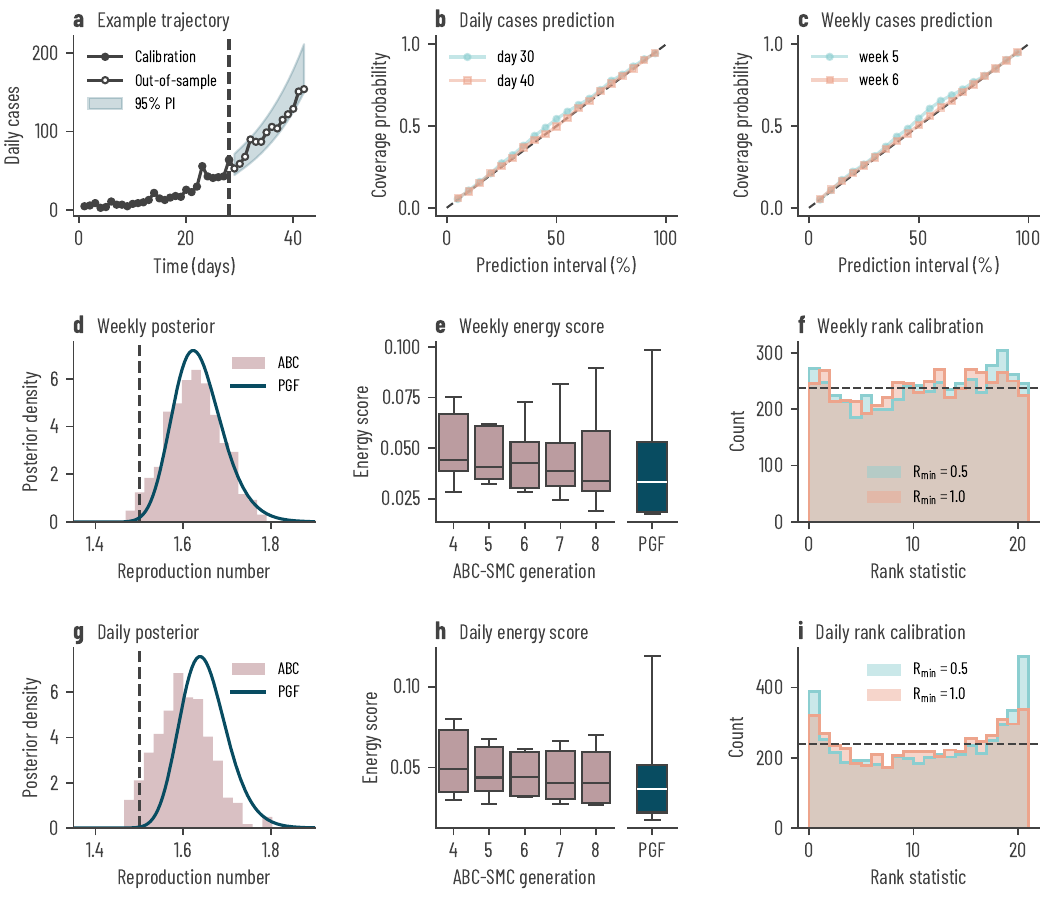}
    \caption{\textbf{Posterior inference on the reproduction number from normal likelihood.} Synthetic data, prediction, and inference settings are identical as in Fig.~\ref{fig:synthetic_experiments}, but we use moment matching with a normal distribution for the likelihood. \textbf{(d-f)} use weekly case counts, \textbf{(g-i)} use daily case counts.}
    \label{fig:synthetic_experiments_normal}
\end{figure}

In Fig.~\ref{fig:synthetic_experiments_sensitivity}, we vary the location parameter $u$ for the translated multivariate lognormal distribution and measure the energy score as in Fig.~\ref{fig:synthetic_experiments}e. The median energy score is minimized around $u = -10$ for both daily and weekly case counts as observables, indicating best performance for inference.
Figure~\ref{fig:synthetic_experiments_daily} reproduce Fig.~\ref{fig:synthetic_experiments}d-f, but we use daily case counts as observables instead of weekly case counts. Results are similar, but we the rank statistics are less uniform, with a larger spike near the upper boundary, indicating the approximate posterior does not perfectly capture the true posterior density and the effects are heighten when predicting low case counts.
Finally, Fig.~\ref{fig:synthetic_experiments_normal} reproduces the results of Fig.~\ref{fig:synthetic_experiments}, but using the normal distribution for moment matching. The approximate posterior deviates more significantly from the ABC-SMC posterior in Fig.~\ref{fig:synthetic_experiments_normal}g when considering daily counts, and we also see larger spikes on both the lower and upper boundaries in Fig.~\ref{fig:synthetic_experiments_normal}i, indicative of further posterior discrepancy.
This suggests that a promising avenue to improve inference is to develop more faithful extraction of the likelihood from PGFs.

\section{Influenza hospital admission model}
\label{sec:influenza_model}

We extend the age-structured SLIR model with asymptomatic transmission and two post-infectious stages, namely $s \in \lbrace \ell, i, \tilde{\imath}, h, H \rbrace$---latent, symptomatic infectious, asymptomatic infectious, pre-hospitalized, and hospitalized.
Latent individuals transition to the infectious stage with probability $\gamma$ per time step, and are asymptomatic with probability $p$; infectious individuals transition to the next stage with probability $\mu$ per time step, independently of their age; symptomatic infectious of age $a$ become pre-hospitalized with probability $q_a$, otherwise they are recovered---all asymptomatic transition to recovered; pre-hospitalized individuals are admitted to the hospital with probability $\delta_a$ per time step; finally, hospitalized individuals recover with probability $\xi$ per time step. $B_{a,a'}$ represents the age-stratified secondary infection per time step for symptomatic infectious and the transmission from asymptomatic infectious is scaled by a factor $\varepsilon$. The offspring PGFs are
\begin{subequations}\label{eq:flu_offspring}
\begin{align}
    G_{t,\ell,a}(\mathbf{x},\mathbf{z}) &= (1-\gamma)\,x_{\ell,a} + \gamma\left[p\,x_{\tilde{\imath},a}z_{\tilde{\imath},a} + (1-p)\,x_{i,a}z_{i,a}\right], \\
    G_{t,\tilde{\imath},a}(\mathbf{x},\mathbf{z}) &= \left[\mu + (1-\mu)\,x_{\tilde{\imath},a}\right] \exp\left[\sum_{a'=1}^{5} \varepsilon\,B_{a,a'}\,(x_{\ell,a'}z_{\ell,a'} - 1)\right], \\
    G_{t,i,a}(\mathbf{x},\mathbf{z}) &= \left[(1-\mu)\,x_{i,a} + \mu\left(1 - q_a + q_a\,x_{h,a}z_{h,a}\right)\right] \exp\left[\sum_{a'=1}^{5} B_{a,a'}\,(x_{\ell,a'}z_{\ell,a'} - 1)\right], \\
    G_{t,h,a}(\mathbf{x},\mathbf{z}) &= (1-\delta_a)\,x_{h,a} + \delta_a\,x_{H,a}\,z_{H,a}, \label{eq:flu_offspring_prehosp}\\
    G_{t,H,a}(\mathbf{x},\mathbf{z}) &= \xi + (1-\xi)\,x_{H,a}.
\end{align}
\end{subequations}
Recovered individuals are not tracked as in the previous SLIR model.
Furthermore, because of the reduced transmission from the asymptomatic individuals, the mean number of secondary infections is rescaled relative to the SLIR model, and the reproduction number reads
\begin{equation}\label{eq:flu_reproduction_number}
    R = \left[1 - p\,(1 - \varepsilon)\right]\frac{\rho(\mathbf{B})}{\mu} \;.
\end{equation}

Each observable $y_\alpha$ accumulates hospital admissions over the duration of a week, and all admissions are assumed to be reported. The contribution PGFs are therefore
\begin{equation}\label{eq:flu_observable_contribution}
    \Xi_{t,H,a}(\mathbf{y}) = y_{\lceil t/7 \rceil} \,,
\end{equation}
for all age group $a$, $\Phi_{t,s,a}(\mathbf{y}) = 1$ for all stages, and $\Xi_{t,s,a}(\mathbf{y}) = 1$ for $s \neq H$.

Finally, the initial cluster is Poisson distributed with mean $\bar{n}_0$, split equally between latent and infectious individuals, and assigned to age group $a$ with probability $w_a$ corresponding to the demographic proportion in the population,
\begin{equation}\label{eq:flu_initial_condition}
    \Psi_0(\mathbf{x}) = \exp\left[\bar{n}_0 \left( \sum_{a=1}^{5} w_a \left( \frac{x_{\ell,a}}{2} + \frac{p\,x_{\tilde{\imath},a} + (1-p)\,x_{i,a}}{2} \right) - 1 \right) \right].
\end{equation}

\begin{table}[h]
    \centering
    \caption{\textbf{Age-dependent parameters of the influenza hospital admission model.} The probability of hospitalization $q_a$ applies to symptomatic infectious individuals. Values are informed from Ref.~\cite{prasad2023multimodeling}.}
    \label{tab:flu_age_params}
    \begin{tabular}{lccccc}
        \toprule
        & [0-4] & [5-19] & [20-49] & [50-64] & [65+] \\
        \midrule
        Hospitalization probability, $q_a$ & 0.0070 & 0.0027 & 0.0056 & 0.0110 & 0.0909 \\
        Mean pre-hospitalized period, $\delta_a^{-1}$ (days) & 1.26 & 1.21 & 1.91 & 2.03 & 1.50 \\
        \bottomrule
    \end{tabular}
\end{table}

\begin{figure}[t]
    \centering
    \includegraphics[width=\linewidth]{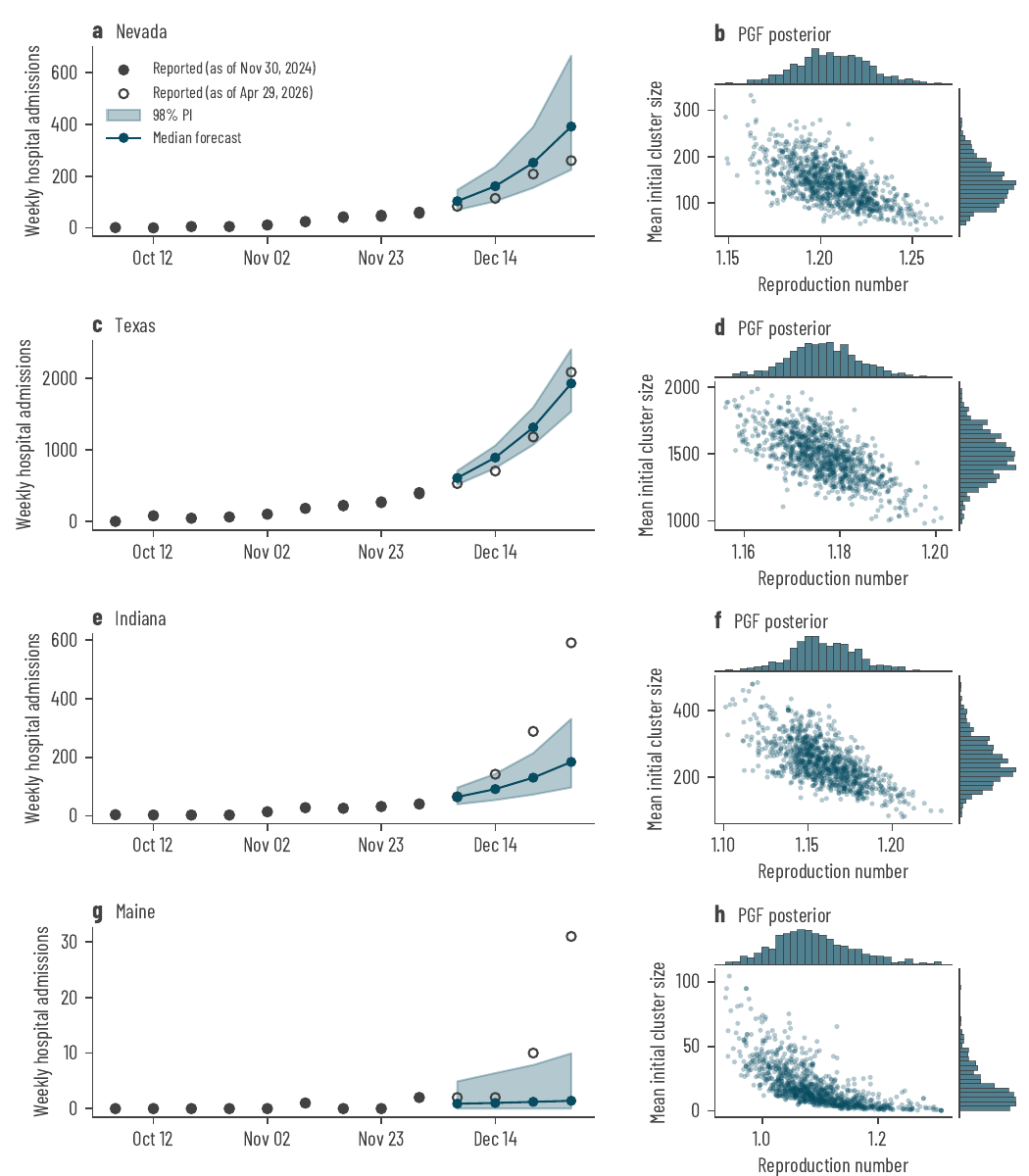}
    \caption{\textbf{Forecasts for additional states.} We use the same model specification as in Fig.~\ref{fig:influenza_hosp_modeling}, but adapt the age-structure to the different states.}
    \label{fig:influenza_hosp_modeling_supp}
\end{figure}

In Fig.~\ref{fig:influenza_hosp_modeling_supp}, we present additional inferences and forecasts for other states. We see that the model perform relatively well when the calibration data is indicative of the future expected growth (Nevada, Texas). However, the model does not capture the epidemic growth when the epidemic onset could not be anticipated from the calibration data alone (Indiana, Maine). Integration of a seasonality factor and additional surveillance data could help capture epidemic onsets.

\section{PGF numerical inversion through fast Fourier transform}
\label{sec:numerical_inversion_fft}

To recover a probability distribution from its generating function, we can leverage the complex plane and Cauchy's integral formula. Let $\Psi(x) = \sum_{n = 0}^\infty P(n)x^n$, then
\begin{align} \label{eq:cauchy_integral}
    P(n) = \frac{1}{n!} \frac{\mathrm{d}^n \Psi}{\mathrm{d} x^n} \Big|_{x=0} = \frac{1}{2\pi \epsilon^{n}}\int_{0}^{2\pi}\Psi(\epsilon e^{\mathrm{i}\omega})e^{-\mathrm{i}\omega n} d\omega.
\end{align}
where $\mathrm{i}$ denotes the imaginary unit, not to be confused with the index for infectious individuals used in previous sections.
The right-hand side is a Fourier integral with tunable parameter $\epsilon$ and can be approximated using the trapezoidal rule,
\begin{align}
    P(n) \approx \frac{1}{N\epsilon^{n}} \sum^{N-1}_{m=0} \Psi( \epsilon e^{2\pi \mathrm{i}m/N})e^{-2\pi \mathrm{i}mn/N} \,,
\end{align}
resulting in a discrete Fourier transform. The absolute error committed by this approximation is
\begin{align}
    E(n)  = \sum^{\infty}_{a=1} P(n + aN)\epsilon^{aN} \,.
\end{align}
Therefore, choosing a large enough $N$ or a suitable $0 < \epsilon < 1$ allows for control over the error.
The approach is directly generalizable in multiple dimensions for a multivariate PGF $\Psi(\mathbf{x})$
\begin{align}
    P(\mathbf{n}) \approx \frac{1}{\prod_{j=1}^d N_j \epsilon_j^{n_j}} \sum_{m_1=0}^{N_1-1} \cdots \sum_{m_d=0}^{N_d-1} \Psi( \boldsymbol{\epsilon} \odot e^{2 \pi \mathrm{i} \mathbf{m} / \mathbf{N}} ) e^{-2\pi \mathrm{i} \mathbf{n} \cdot \mathbf{m}/\mathbf{N}} \,.
\end{align}

\end{document}